\documentclass[10pt,conference]{IEEEtran}

\usepackage[T1]{fontenc}
\usepackage[utf8]{inputenc}

\usepackage{subfiles}                  
\usepackage{comment}                   
\usepackage{xcolor}                    
\usepackage{eurosym}                   
\usepackage{gensymb}                   
\usepackage{bbding}                    
\usepackage{soul}                      
\usepackage{url}                        
\usepackage{hyperref}                   

\usepackage[numbers,sort&compress]{natbib}

\usepackage{graphicx}                  
\usepackage[caption=false]{subfig}     
\usepackage{wrapfig}                   
\usepackage{array}                     
\usepackage{multirow}                  
\usepackage{booktabs}                  
\usepackage{tabularx}                  
\usepackage{longtable}                 
\usepackage{colortbl}                  
\usepackage{rotating}                  
\usepackage{float}                     
\usepackage{csvsimple}                 

\usepackage{amsmath}
\usepackage{amssymb}
\usepackage{amsfonts}
\usepackage{mathtools}

\usepackage[ruled]{algorithm2e}
\usepackage{minted}

\usepackage[most]{tcolorbox}

\definecolor[named]{ACMBlue}{cmyk}{1,0.1,0,0.1}
\definecolor[named]{ACMYellow}{cmyk}{0,0.16,1,0}
\definecolor[named]{ACMOrange}{cmyk}{0,0.42,1,0.01}
\definecolor[named]{ACMRed}{cmyk}{0,0.90,0.86,0}
\definecolor[named]{ACMLightBlue}{cmyk}{0.49,0.01,0,0}
\definecolor[named]{ACMGreen}{cmyk}{0.20,0,1,0.19}
\definecolor[named]{ACMPurple}{cmyk}{0.55,1,0,0.15}
\definecolor[named]{ACMDarkBlue}{cmyk}{1,0.58,0,0.21}
\hypersetup{
    colorlinks=true,
    citecolor=ACMPurple,    
    linkcolor=ACMBlue,      
    urlcolor=ACMDarkBlue    
}

\newtcolorbox{researchqbox}{
  enhanced,
  colback=blue!4!white,
  colframe=blue!60!black,
  boxrule=0.6pt,
  arc=2mm,
  left=6pt,
  right=6pt,
  top=4pt,
  bottom=4pt,
  drop shadow,
  breakable,
}

\begin{document}

\title{Tiny Machine-Learning Operations within\\ Cyber-Physical Systems: a Field Study}

\author{%
  \IEEEauthorblockN{Filippo Scaramuzza}
  \IEEEauthorblockA{\textit{Eindhoven University of Technology}\\
  \textit{Tilburg University}\\
  f.scaramuzza@tue.nl}
  \and
  \IEEEauthorblockN{Damian~A.~Tamburri}
  \IEEEauthorblockA{\textit{University of Sannio}\\
  \textit{JADS / NXP Semiconductors}\\
  d.a.tamburri@tue.nl}
}

\maketitle

\begin{abstract}
Machine-Learning Operations (MLOps) is maturing into a software-engineering
discipline, yet its tiny-scale variant (TinyMLOps), targeting the
resource-constrained microcontrollers embedded in cyber-physical systems
(CPS), remains poorly understood in industrial practice. Opaque models, noisy
heterogeneous data, and tight memory budgets hinder adoption in safety-critical
settings, where most decisions still rely on human experts. We report a field
study of an end-to-end, knowledge-centred TinyMLOps pipeline that fuses domain
physics, expert speculation, and sensor streams to deliver explainable,
low-footprint models deployable on-device. The pipeline spans automated
collection and cleaning of heterogeneous time series, knowledge-driven feature
construction, interpretable regularised models, and rolling temporal
cross-validation under concept drift. We evaluate it on 4.4 GB of data from two
offshore-wind cable-trenching campaigns. The classifier anticipates harmful
load peaks up to three minutes ahead at 0.84 AUC within a 32 kB footprint on an
ARM Cortex-M4; an ablation shows that injecting prior knowledge halves false
alarms and surfaces actionable operational rules. Replaying recommendations in
operational dashboards indicates an 11\% reduction in non-productive time. We
distil engineering lessons and validity threats for trustworthy TinyMLOps in
CPS, and release our open-source pipeline and declarative feature encodings to
support reproducibility.

\end{abstract}


\section{Introduction}\label{sec:introduction}\label{ch:introduction}
\IEEEPARstart{M}{achine-Learning} Operations (MLOps) has emerged as the
software-engineering discipline that turns experimental machine-learning (ML)
artefacts into dependable, maintainable, and continuously delivered production
systems~\cite{kreuzberger2023machine,tamburri2020sustainable}. As ML moves from
the data centre to the physical edge, a \emph{tiny} counterpart, TinyMLOps, is
taking shape: the engineering of ML models small enough to run on the
microcontrollers embedded in everyday devices~\cite{tinyml,faubel2023mlops}.
Nowhere is this shift more consequential than in Cyber-Physical Systems (CPS),
where computation, networking, and physical processes are tightly coupled through
feedback loops~\cite{lee2007computing}. CPS extend the Internet of Things from
sensing and transmission to \emph{control}~\cite{Nandhini2022ARO}, and they are
the technical nexus of Industry~4.0 (I4.0), whose market is projected to grow
several-fold within the decade~\cite{fbi-i4,fmi-cps}. Operating ML \emph{inside}
such systems raises software-engineering questions
that classic, cloud-centric MLOps does not answer: how to engineer pipelines that
respect kilobyte-scale memory budgets, remain explainable enough for
safety-critical sign-off, and stay trustworthy as the physical environment
drifts.

Despite this momentum, the convergence of ML and CPS in industry is still largely
aspirational. Most deployed CPS depend on heavy human monitoring and reactive
diagnosis; the few ML models in production are often unavailable for complex
assets, or incomplete and inaccurate, injecting uncertainty that erodes the
reliability and quality of operations~\cite{asmat2022uncertainty}. The barrier is
as much a software-engineering problem as a modelling one: empirical-risk
minimisation optimises a loss over sampled inputs and offers no guarantee on the
inputs encountered operationally~\cite{gu2019towards}, while opaque models resist
the assurance arguments that safety-critical CPS require~\cite{gharib2018safety}.
A growing consensus therefore advocates \emph{hybrid} approaches that combine
physics and data, embedding pre-existing scientific knowledge into ML to improve
both transparency and performance~\cite{von2021informed,beckh2021explainable,
karniadakis2021physics}. Realising such approaches on-device, under an
engineering process that can be operated and maintained over time, is precisely
the TinyMLOps challenge this paper addresses.

We study this challenge \emph{in the field} of ocean engineering, an exemplary,
high-stakes setting where CPS such as Remotely Operated Vehicles (ROVs) perform
safety-critical subsea power-cable burial under pervasive uncertainty from
variable soil, equipment, and human control~\cite{panda2023machine,DIAZ2020107381}.
The domain relies heavily on expert, model-based control where errors are
expensive and occasionally life-threatening, yet its assets continuously emit
rich, largely unexploited sensor streams. It thus offers a representative,
demanding test of whether disciplined data engineering and physics-informed bias
can convert noisy CPS streams into trustworthy on-device decision support.

This paper presents a \emph{field study} of the engineering,
deployment, and empirical evaluation of an end-to-end, knowledge-centred
TinyMLOps pipeline, grounded in two large-scale offshore-wind campaigns.
Concretely, we make the following contributions:
\begin{itemize}
  \item A \emph{reusable TinyMLOps pipeline architecture} that fuses domain
    physics, expert speculation, and heterogeneous sensor streams into
    explainable, sub-32\,kB models deployable on an ARM Cortex-M4, with rolling
    temporal cross-validation for robust estimation under concept drift
    (Sections~\ref{ch:design} and~\ref{ch:projectmethodology}).
  \item A \emph{knowledge-engineering method} that encodes expert speculations as
    configurable, lagged feature transformations and empirically validates or
    refutes them against data, turning tacit operator know-how into auditable
    model evidence (Sections~\ref{ch:projectmethodology}
    and~\ref{ch:results}).
  \item An \emph{empirical evaluation} on 4.4\,GB of data from the Kaskasi~II
    (42 inter-array cables) and South Fork Wind (11 cables) campaigns, showing
    that the classifier anticipates harmful load peaks up to three minutes ahead
    at 0.84 AUC, that prior-knowledge injection halves false alarms, and that
    replayed recommendations would cut non-productive time by 11\%
    (Section~\ref{ch:results}).
  \item A distillation of \emph{engineering lessons and threats to validity} for
    building trustworthy TinyMLOps in CPS, together with an open-source release
    of our pipeline, declarative feature encodings, and supplementary evaluation
    artefacts to foster reproducible knowledge engineering in
    resource-constrained CPS (Sections~\ref{ch:discussion}--\ref{sec:replication}).
\end{itemize}

\section{Related Work}\label{ch:literatureReview}
\label{bg}
We position our study at the intersection of four research strands whose
convergence in safety-critical CPS remains under-studied.

\subsection{From Predictive Control to Learning-Based Control}
\label{ch:literatureReview:sec:predictiveControl}
The control of CPS has evolved through four broad stages that progressively
admit data and learning~\cite{prag2022toward,hewing2020learning}.
\emph{Model Predictive Control} (MPC) uses an explicit model to choose actions
by online optimisation and excels in well-understood
systems~\cite{sun2019resilient}. \emph{Data-Driven MPC} relaxes the assumption
of a perfect model by fitting plant dynamics from historical (offline) data,
marking the entry of ML into the loop. \emph{Data-Driven Controller Tuning}
then adapts controller parameters online from data, and finally
\emph{Learning-Based Data-Driven Control} drops the explicit physical model
altogether, iteratively learning a control law that copes with disturbance and
parameter drift~\cite{alippi2016model}. While this trajectory trades explicit
physics for data, the model-free extreme sacrifices the very
interpretability and assurance that safety-critical CPS demand, motivating the
\emph{hybrid} stance we adopt. Crucially, this literature concentrates on
control-theoretic performance and rarely treats the resulting learning component
as a software artefact that must be built, deployed, operated, and maintained.

\subsection{MLOps and TinyML as Software Engineering}
\label{ch:literatureReview:sec:mlops}
That software-engineering gap is precisely what MLOps addresses. Kreuzberger
\emph{et al.}~\cite{kreuzberger2023machine} consolidate MLOps into a set of
principles, components, and roles for operationalising ML, observing that many
industrial ML projects fail not at modelling but at \emph{productionisation}.
Tamburri~\cite{tamburri2020sustainable} frames the sustainability and technical
debt of ML-intensive systems, and Kolltveit and Li~\cite{kolltveit2022operationalizing}
identify model \emph{deployment} as one of the most frequently reported obstacles
to ML adoption. Process models such as CRISP-ML(Q)~\cite{studer2021towards}
extend CRISP-DM with explicit quality-assurance and monitoring phases, while
empirical work on continuous integration for ML
systems~\cite{zampetti2022continuous} documents the engineering friction of
keeping such systems alive. For the embedded edge, Faubel \emph{et
al.}~\cite{faubel2023mlops} show that applying MLOps within the heterogeneous,
resource-constrained landscape of I4.0 requires an architecture that spans the
whole CPS and interfaces with legacy tooling, a challenge compounded by kilobyte
memory budgets and diverse hardware platforms. This body of work establishes
\emph{what} good ML engineering looks like, but is dominated by cloud-scale,
data-centre assumptions; \emph{TinyMLOps}~\cite{tinyml} for safety-critical CPS
is comparatively unexplored and almost never reported through industrial field
studies.

\subsection{Safety, Uncertainty, and Explainability in CPS}
\label{sec:literatureReview_predictiveControlCps}
Two forces pull ML into CPS architectures: the abundance of operational data,
and ML's ability to tackle control problems that resist conventional algorithmic
solutions~\cite{pereira2020challenges}. Yet CPS are typically
safety-critical~\cite{gharib2018safety}: failures can cause injury or major
asset damage, so safety must be assured before deployment. ML complicates
assurance because its inductive predictions inherently carry a probability of
error~\cite{gu2019towards}, and because epistemic uncertainty arises from human
behaviour, natural processes, and the technology stack
itself~\cite{asmat2022uncertainty}. The dominant mitigation in recent literature
is to \emph{inform} ML with prior knowledge: physics-informed
learning~\cite{karniadakis2021physics}, taxonomies of informed
ML~\cite{von2021informed}, and pipelines that foreground
explainability~\cite{beckh2021explainable} all argue that domain knowledge
improves transparency and can boost accuracy. We operationalise this stance as
an engineering method, encoding expert speculation as auditable, testable
features rather than a purely algorithmic device.

\subsection{Machine Learning in Ocean Engineering}
Interest in applying ML to ocean engineering is rising
sharply~\cite{DIAZ2020107381,panda2023machine}, spanning autonomous-vehicle
control and environmental modelling. Subsea cable installation, however, remains
governed by expert, model-based assessments and real-time descriptive
dashboards; predictive, data-driven control of the trenching process is scarce,
and the few models in use (e.g.\ static burial-assessment
models~\cite{warringa2019modellling}) struggle to track the dynamic
configurations and soil conditions of modern campaigns. The domain thus
exemplifies the broader CPS adoption gap: abundant data, high stakes, and an
entrenched reliance on tacit expertise.

\begin{researchqbox}
\noindent\textbf{Synthesis and Gap.}
Existing work addresses these concerns largely in isolation. Learning-based
control prioritises control-theoretic performance over software lifecycle
engineering, MLOps research centres on cloud-scale infrastructure rather than
kilobyte-constrained embedded targets, and informed ML typically treats domain
knowledge as a fixed modelling constraint instead of a testable engineering
artefact. Consequently, industrial field evidence on how to build, validate, and
operate explainable TinyML pipelines in safety-critical CPS remains scarce.
\end{researchqbox}

\section{Empirical Study Design}\label{ch:design}
\subsection{Research Questions}\label{sec:design:rqs}
The study is driven by three research questions (RQs) progressing from
\emph{engineering} the pipeline (Sections~\ref{ch:useCase}--\ref{ch:projectmethodology}),
to its \emph{empirical efficacy} (Sections~\ref{sec:results:labels}--\ref{sec:results:performance}),
to its \emph{generalisability} (Section~\ref{ch:results:southfork}). Throughout,
\textit{high uncertainty} denotes the unpredictable factors affecting CPS
behaviour, and \textit{performance} denotes the safe improvement of productivity
or quality.

\begin{researchqbox}
\noindent\textbf{RQ1 (Engineering).} How can an explainable, knowledge-centred
TinyMLOps pipeline be designed to enhance the performance of a Cyber-Physical
System operating under high uncertainty, within embedded resource constraints?
\end{researchqbox}

\begin{researchqbox}
\noindent\textbf{RQ2 (Efficacy).} To what extent can the explainable pipeline
empirically validate or refute expert speculation and augment operational
decision-making in the CPS?
\end{researchqbox}

\begin{researchqbox}
\noindent\textbf{RQ3 (Generalisability).} How well does the approach transfer to
an out-of-sample campaign with materially different operating conditions?
\end{researchqbox}

\subsection{Research Method}\label{sec:design:method}
We adopt an in-depth, \emph{instrumental} case-study
method~\cite{harling2012overview}: the case (subsea cable trenching by an ROV)
is studied not for its own sake but as an instrument to understand the broader
phenomenon of integrating explainable TinyML into safety-critical CPS under
uncertainty. Following Harling, the study is bounded (specific trenching
projects within one industrial partner), set in its natural operational context,
and pursued holistically through multiple evidence sources: internal records,
expert elicitation, and high-frequency sensor logs. The case-study method suits
a contemporary phenomenon examined in its real setting, where the boundary
between phenomenon and context is blurred and controlled experimentation is
infeasible.

To support rigour and reviewer assessment we align the design with the ACM
SIGSOFT empirical standard for case studies: we state the unit of analysis and
case boundaries, triangulate data sources, describe the context in detail, and
maintain a chain of evidence from raw sensor streams to fitted coefficients and
operational recommendations. Validity is discussed in
Section~\ref{ch:threats} along the construct, internal, external, and conclusion
dimensions.

\subsection{Case and Context}\label{sec:design:context}
The unit of analysis is a single trenching pass of one inter-array cable, and
the embedded decision-support model that runs alongside it. The primary case is
the Kaskasi~II offshore-wind campaign (42 inter-array cables, predominantly
sandy soil); the out-of-sample case is South Fork Wind (11 cables, harder and
more variable soil). Section~\ref{ch:useCase} details the equipment, the
trenching process, and the prevailing control practice that the pipeline augments.

\subsection{Data Collection}\label{sec:design:dataCollection}\label{sec:projectMethodology:subsec:dataCollection}\label{ch:methodology:sec:dataPreparation}
Four primary data sources are used, collected through the \texttt{trencher} Python package and SQL scripts:
(1)~\emph{Daily Progress Reports} (DPRs), (2)~\emph{Settings}, (3)~\emph{Sensor
Data}, and (4)~\emph{Soil-Layer Data}. DPRs are retrieved from an Azure SQL
time-registration database as CSV, pre-filtered to the trenching
vessel, activity, first pass, and relevant projects.
Settings and sensor data follow a single flow: onboard PLCs on the trencher feed
a connector on the vessel, which uploads to an on-board historical sensor
database synced to an office twin, both exposing a web API. Trencher tags
are discovered through a wildcard \texttt{POST} call, then retrieved per cable
by tag-name filter and DPR start/end UTC timestamps and joined on
timestamp into per-cable CSV tables to speed cross-validation.
Soil-layer data is queried from an Azure-hosted PostgreSQL aggregated database
into a single CSV for all cables.

Although many projects used the same trencher, the study concentrates on
Kaskasi~II for three engineering reasons: it was among the first to incorporate
the Aft Cable Guide (Section~\ref{ch:useCase}), it has richer and more reliable
instrumentation, and it post-dates the deployment of operational dashboards. Its
mostly sandy soil yields consistent processes with fewer confounders (operator
interventions, boulders), and its varied-length inter-array cables form a
representative sample for assessing cross-cable uncertainty. Only first-pass
trenching is modelled, as subsequent passes are rare and highly idiosyncratic.

\subsection{Analysis Procedure}\label{sec:design:analysis}
Three complementary techniques answer the RQs. First, to estimate predictive
performance honestly under \emph{concept drift}, we use \emph{rolling temporal
cross-validation}: models are trained on cables seen so far and evaluated on the
next, never using future data, with standardisation statistics derived only from
past cables to avoid leakage. Second, to test expert knowledge (RQ2) we perform
\emph{coefficient-stability analysis}: each speculation is encoded as a
configurable feature transformation, and we inspect the sign and stability of its
fitted coefficient, both per cable and cumulatively across cables, to
decide whether it is empirically supported and may enter the knowledge base.
Third, we run an \emph{ablation} contrasting the knowledge-informed pipeline with
a knowledge-free baseline. Predictive quality is reported via Area Under the
ROC Curve (AUC) and false-alarm rate; operational benefit is estimated by
replaying model recommendations against recorded campaigns to measure the
implied reduction in non-productive time.

\section{Field Study Context: Subsea Trenching and TinyMLOps}\label{ch:useCase}
\subsection{Subsea Cable Burial}
\begin{figure*}
    \centering
    \includegraphics[width=\textwidth]{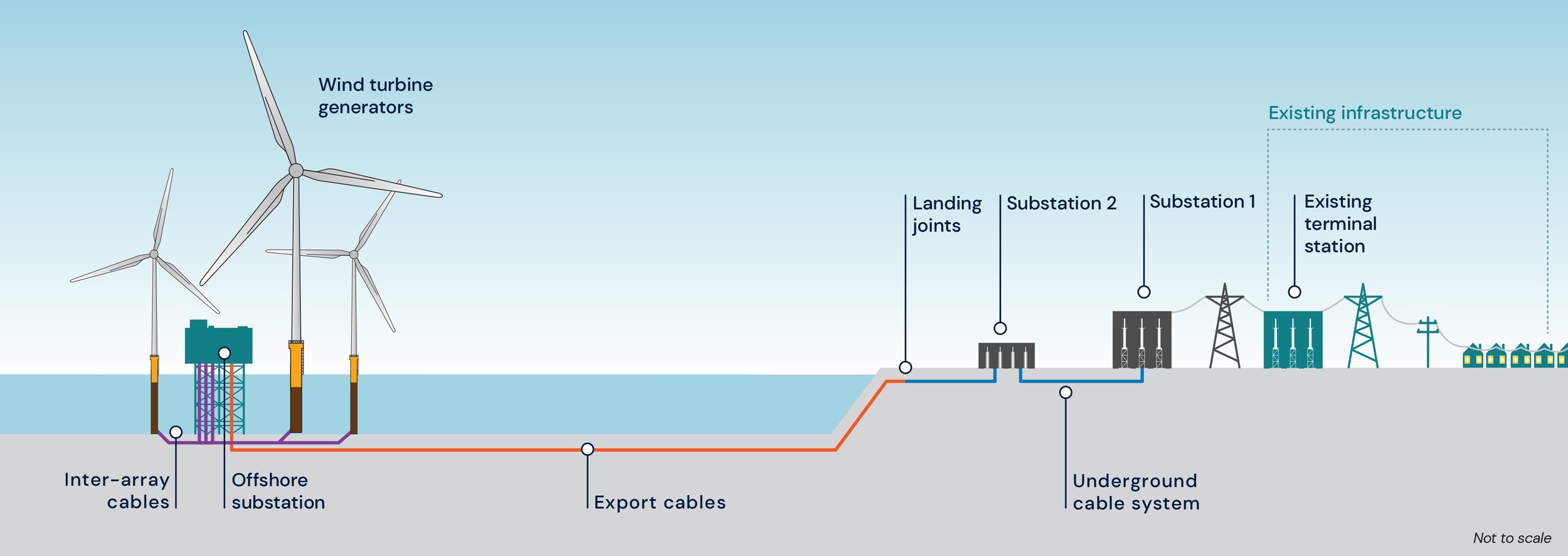}
    \caption{Offshore wind-farm overview~\cite{starofthesouth}.}
    \label{fig:OWF_overview}
\end{figure*}

\begin{figure}
    \centering
    \includegraphics[width=0.42\textwidth]{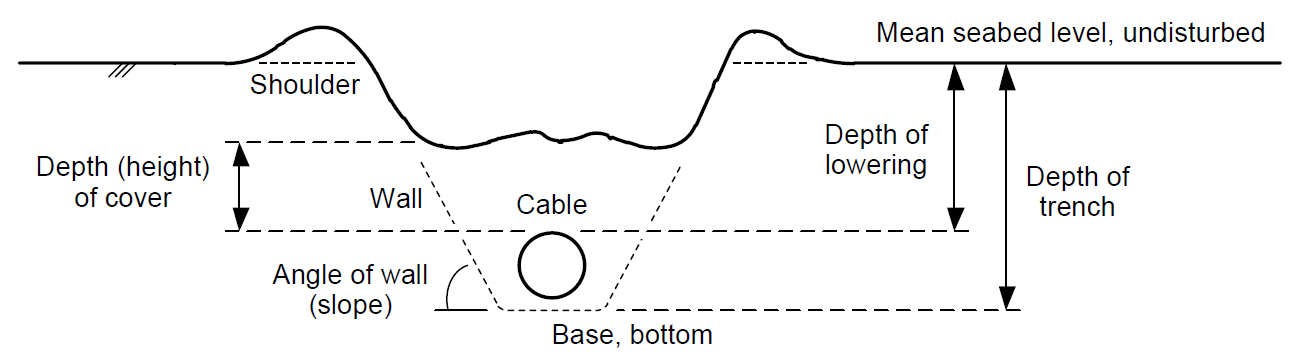}
    \caption{Cable-burial design and Depth of Lowering (DoL).}
    \label{fig:depthOfLowering}
\end{figure}

Cable installation is one of the most intricate tasks in offshore-wind
construction~\cite{GONZALEZRODRIGUEZ201710}, and the rapid expansion of wind
farms has sharply increased demand for cable deployment. Two cable types matter
for an Offshore Wind Farm (OWF, Figure~\ref{fig:OWF_overview}): export cables
between shore and the offshore site, and inter-array cables (IACs) laid between
turbines or to an Offshore Substation (OSS).
Burial protects the cable and its environment; its key qualitative target is the
Depth of Lowering (DoL), the distance from the reference seabed to the top of the
cable (Figure~\ref{fig:depthOfLowering}).
Of the four burial techniques classified by Kraus and
Carter~\cite{KRAUS2018251} (ploughing, jetting, mechanical cutting, and
horizontal directional drilling), this study concerns \emph{jetting}, in which
high-speed water jets fluidise the soil so a pre-laid cable sinks under its own
weight~\cite{jmse8060460}, with \emph{cutting} available for harder
substrates~\cite{pyrah2010cable}.

\subsection{The Cyber-Physical System: a Trenching ROV}
\begin{figure}
    \centering
    \includegraphics[width=0.46\textwidth]{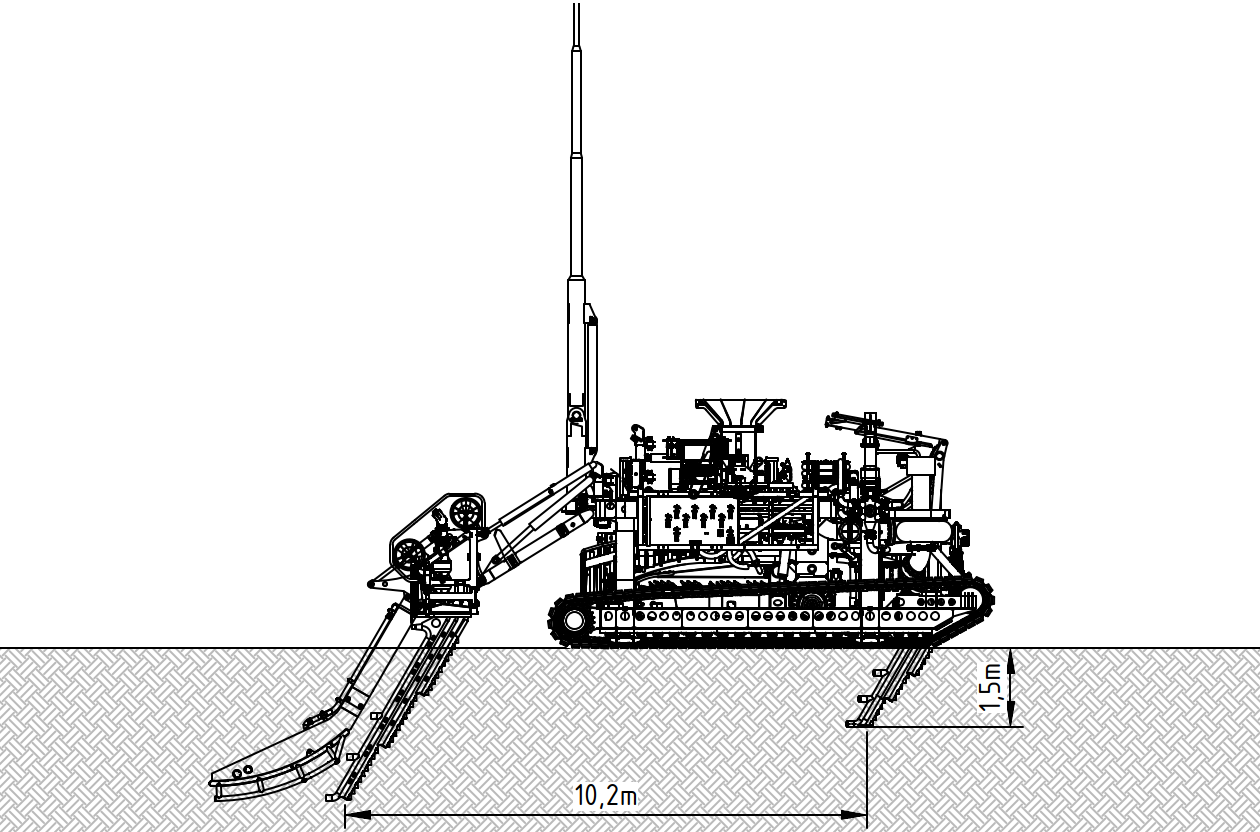}
    \caption{Technical drawing of the trenching ROV, highlighting the FWD/AFT jetting tools and the AFT Cable Guide (ACG).}
    \label{fig:cbt1100Trencher}
\end{figure}

The industrial partner operates a specialised trenching ROV (the CBT1100),
remotely managed by operators aboard a multi-purpose vessel. The trencher
carries forward (FWD) and aft (AFT) jetting swords, a forward cutting chain for
resilient soils, and, critically, an \emph{AFT Cable Guide} (ACG). Whereas a
cable normally sinks to depth once the soil is fluidised, the ACG actively
\emph{guides} (pushes) the cable to the required depth. With the ACG in place,
DoL becomes a secondary concern; the binding constraint shifts to the
\emph{resistance encountered by the ACG}. Figure~\ref{fig:cbt1100Trencher} shows
the antenna, the FWD/AFT jetting tools, and the ACG.
The primary case, the Kaskasi~II OWF, comprised 32 monopiles, 39 transition
pieces, the burial of 42 IACs, and 38 wind-turbine generators, for 342\,MW total
output, fully operational since early 2023~\cite{4coffshore}, and was executed
entirely in jetting mode.

\subsection{Prediction Target: Harmful Cable-Guide Load Peaks}
\label{sec:caseStudy:target}
Productivity (trenching speed) is routinely sacrificed to safety. Increasing
speed reduces the time available to form the trench, which can (i) overburden the
jetting/cutting instruments and (ii) raise the load on the ACG. In normal
operation the ACG force tracks the cable force; if the cable force exceeds a
safety limit, however, the ACG retracts to avoid cable damage, decreasing DoL
(possibly below requirement) and necessitating a costly second pass over a
partially buried cable. The engineering target of our pipeline is therefore the
\emph{anticipation of harmful peaks in ACG load}, far enough ahead for the
operator (or an automated controller) to react. Measurable influences fall into
\emph{directly controllable} factors (power to tracks, jetting-nozzle pressure,
cutting power, instrument depth), \emph{indirectly controllable} factors
(instrument loads that foreshadow ACG load), and \emph{uncontrollable} factors
(soil composition, density, cohesiveness).

\subsection{Current Predictive Control}\label{sec:caseStudy:control}
Two mechanisms govern trenching today. Before a project, a Burial Assessment
Study (BAS) by geotechnical engineers sets an initial, conservative trenching
speed from a static, physics-informed model~\cite{warringa2019modellling}.
During operations, engineers consult a Grafana time-series dashboard, fed by the
same 1\,Hz sensor streams with up to a 15-minute office latency, to adjust
speed. In the categories of Prag \emph{et
al.}~\cite{prag2022toward}, the BAS is open-loop, model-based, offline, and
fixed, whereas the dashboard approach is closed-loop, model-free, online, and
adaptive but \emph{descriptive} only. Both leave the explanatory strength of
load-peak causes empirically unverified, encouraging over-cautious speeds: a
recorded export-cable comparison found actual productivity exceeding BAS
predictions by 56\% at 1.5\,m DoL. This is the opening our pipeline targets.

\subsection{Expert Speculations on Cable-Guide Load}\label{sec:caseStudy:speculations}
Operators and engineers hold tacit, sometimes conflicting beliefs about what
drives ACG-load peaks. Elicited from four domain experts, these
\emph{speculations} (Table~\ref{tab:speculationsAcgLoad}) are the raw material
our knowledge-engineering method encodes and tests (Section~\ref{ch:results}):
each is a hypothesis the pipeline can confirm, refine, or refute against data.
\begin{table}
\centering
\footnotesize
\setlength{\tabcolsep}{4pt}
\caption{Elicited expert speculations on causes of ACG-load peaks.}
\begin{tabular}{@{}p{0.45cm}p{2.4cm}p{4.25cm}@{}}
\toprule
No. & Cause & Hypothesised effect on ACG load \\
\midrule
1 & Steering movements & Increase: ACG scrapes the trench side \\
2 & Increasing Depth of Lowering & Increase: ACG pushes cable deeper \\
3 & Increasing trencher speed & Increase: less trenching time \\
4 & Consistency of instrument loads & Stable load under stable conditions \\
5 & Soil density \& cohesiveness & Increase: more soil resistance \\
6 & Reduced sword depth & Increase: less pre-trenching depth \\
7 & Presence of boulders & Increase: sudden elevated resistance \\
\bottomrule
\end{tabular}
\label{tab:speculationsAcgLoad}
\end{table}

\section{Data Engineering Perspective}\label{ch:data}
\subsection{Sources, Scale, and Governance Gaps}
Across the four collected sources (4.4\,GB total;
Section~\ref{sec:design:dataCollection}), sensor data streams from the trencher
at 1\,Hz across roughly 650 virtual-sensor \emph{tags}, each keyed by source and
measurement type (track speeds, jetting pressures and widths, cable-guide load
and depth, navigation). A first governance problem is \emph{semantic drift}: the
data inconsistently mixes ``depth'' and ``elevation'' references, so we normalise
all vertical references to four explicit quantities: depth under seabed ($DUS$)
and under water ($DUW$), and elevation from seabed ($EFS$) and from water
($EFW$).

A second problem is \emph{coarse, unreliable temporal labelling}. DPR intervals
are too wide to delimit active trenching: for one cable the DPR spanned nearly
three hours, whereas the sensor traces show trenching beginning many minutes
later and the kilometre-point (KP) increasing only thereafter. We therefore
derive the active window directly from the data (the longest strictly-monotonic
KP segment) rather than trusting the DPRs, and identify cables by KP (a
physically meaningful key) instead of timestamps. These exemplify the
instruction-validity and configuration-mismatch frictions that erode the
reproducibility of operational ML.

\subsection{Descriptive Insights Driving Design}
\begin{figure}
    \centering
    \includegraphics[width=.46\textwidth]{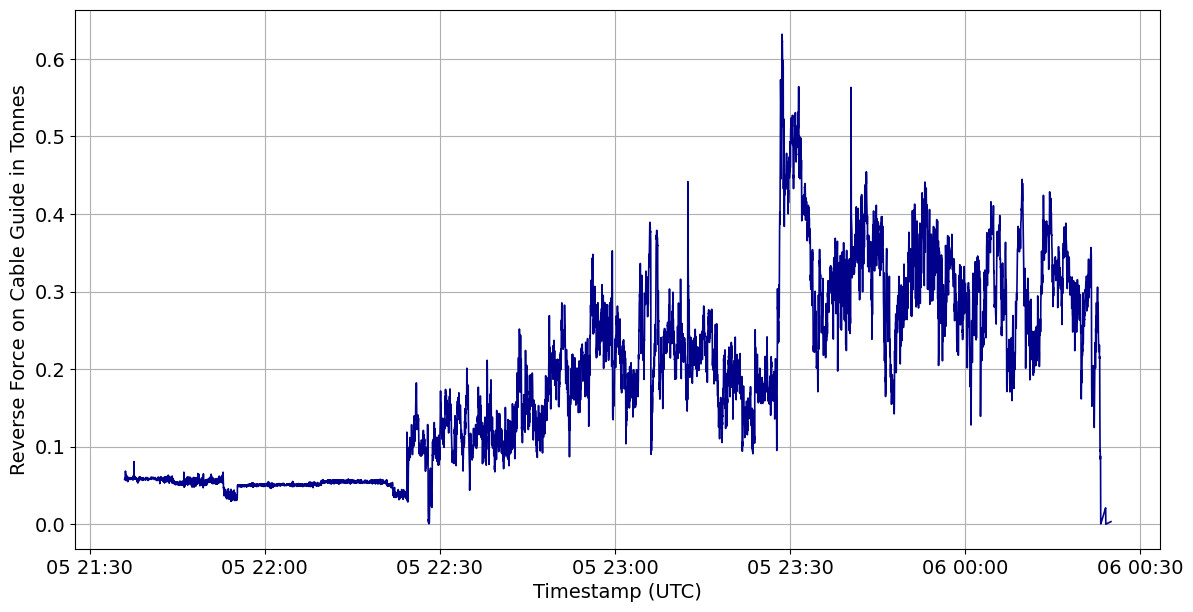}
    \caption{ACG load over time for a representative cable: sudden surges (peaks), not steady level, are the safety-relevant signal.}
    \label{fig:descriptiveAnalysisLoadcellOverTime}
\end{figure}

Sensor measurements group into three operationally meaningful families used live
on the dashboard: track, jetting, and cable-guide measurements. Two descriptive
findings shaped the pipeline. First, the prediction signal is \emph{peaks}, not
level: the ACG load exhibits sudden force surges
(Figure~\ref{fig:descriptiveAnalysisLoadcellOverTime}) that matter far more than
its steady value, motivating the peak \emph{labelling} of
Section~\ref{subsec:loadPeakLabel} rather than regression on raw load. Second,
effects are \emph{delayed by geometry}: at a typical 250\,m/h, the 1\,Hz stream
yields $\approx$14 observations per metre, and because the forward tools sit
$\sim$15\,m ahead of the ACG, their influence reaches the cable guide only after
$\approx$3.6\,minutes (Eq.~\ref{eq:data:lead}), providing the physical basis for
both the lagged features and the three-minute anticipation horizon.
\begin{equation}
\label{eq:data:lead}
250\,\text{m/h}=4.17\,\text{m/min}\;\Rightarrow\;\frac{15\,\text{m}}{4.17\,\text{m/min}}\approx 3.6\,\text{min}.
\end{equation}
Soil-layer data provides upper layer boundaries ordered by relative density or
cohesiveness, of which only those intersecting the ACG ($\sim$1.5\,m depth) and
jet depth matter, motivating the soil-proportion features in
Section~\ref{subsec:dataConstruction}. Representative soil profiles and the
descriptive-analysis scripts for per-signal traces are provided in the
replication package.

\section{TinyML Modelling and Ops Pipeline}\label{ch:projectmethodology}
\subsection{Pipeline Architecture and Operations}\label{sec:pipeline:arch}
\begin{figure}
    \centering
    \includegraphics[width=.46\textwidth]{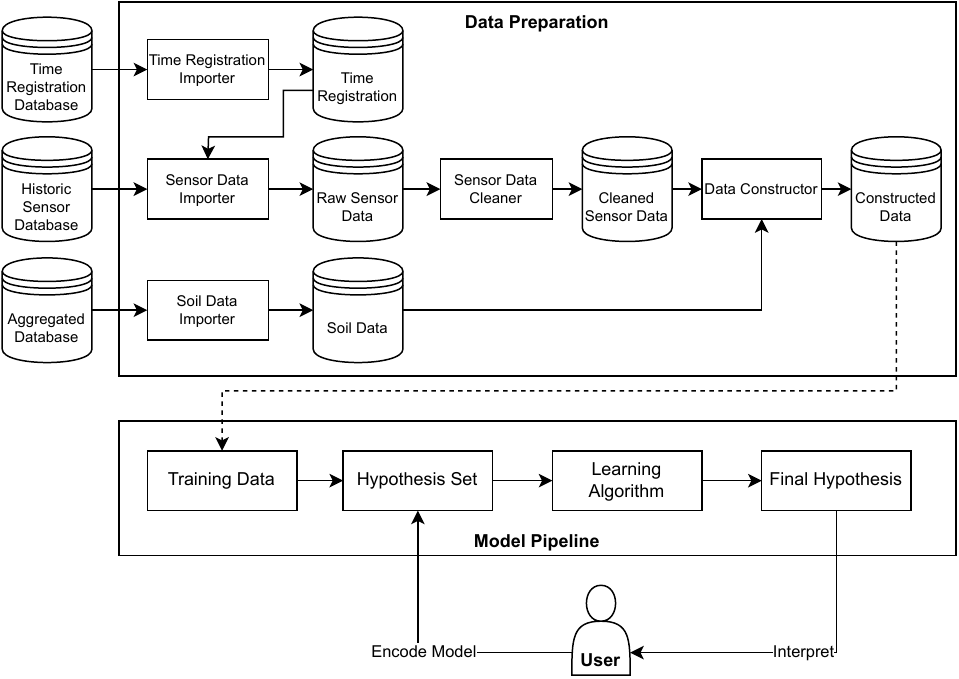}
    \caption{Implemented TinyMLOps pipeline and user interaction.}
    \label{fig:implementedPipeline}
\end{figure}

To keep the system maintainable and the knowledge-engineering loop fast, data
construction is separated from training: the former is comparatively expensive
and materialised once per cable, while the latter is cheap and re-run as
experts vary the encoded knowledge. Concretely, the pipeline is a configurable,
object-based component (implemented in Python, with on-device inference in
MicroPython\footnote{\url{https://micropython.org/}}) exposing the operations:
\emph{read cable} (load a cable, strip the first/last 200 manoeuvring
observations); \emph{aggregate cable} (construct features by applying a chosen
aggregation over a user-specified, possibly lagged window); \emph{train} and
\emph{predict} (fit/apply an interpretable linear or logistic model, optionally
$\ell_1$-regularised); \emph{plot/return coefficients} (per-cable and cumulative,
for empirical knowledge verification); and \emph{loop cables} (rolling temporal
cross-validation across cables).
Figure~\ref{fig:implementedPipeline} summarises the data-construction and model
pipeline together with the points of user interaction.
A model is fully described by a YAML specification; editing it reconfigures the
features without touching code, making \emph{knowledge exploration}
practical. Each entry follows the schema in Table~\ref{tab:modelEncoding}: a
source column, an aggregation (immediate difference \texttt{diff}, windowed mean
difference \texttt{avgdiff}, mean, or variance), a window $(-a,b)$ spanning
$[t-a,\,t-a+b]$, and whether to standardise. The same engine serves two modes: an
\emph{exploration} mode that searches over candidate (speculative) encodings, and
a \emph{validation} mode that admits only certified knowledge. The fitted model
occupies a 32\,kB footprint on an ARM Cortex-M4, enabling on-device
recommendations without waiting on the dashboard's office-side latency.
\begin{table}
\caption{Model-encoding schema (one row per feature).}
\centering
\footnotesize
    \begin{tabular}{ll}
    \toprule
    Field            & Allowed values \\
    \midrule
    Column name      & a column of the constructed data \\
    Aggregation      & \{diff, avgdiff, mean, var\} \\
    Window           & (start, length) $=(-a,b)$, i.e.\ $[t-a,\,t-a+b]$ \\
    Standardisation  & \{yes, no\} \\
    \bottomrule
    \end{tabular}
    \label{tab:modelEncoding}
\end{table}

\subsection{Knowledge-Driven Feature Construction}\label{subsec:dataConstruction}
Two mechanisms turn raw, heterogeneous streams into knowledge-bearing features.

\begin{figure}
    \centering
    \includegraphics[width=.46\textwidth]{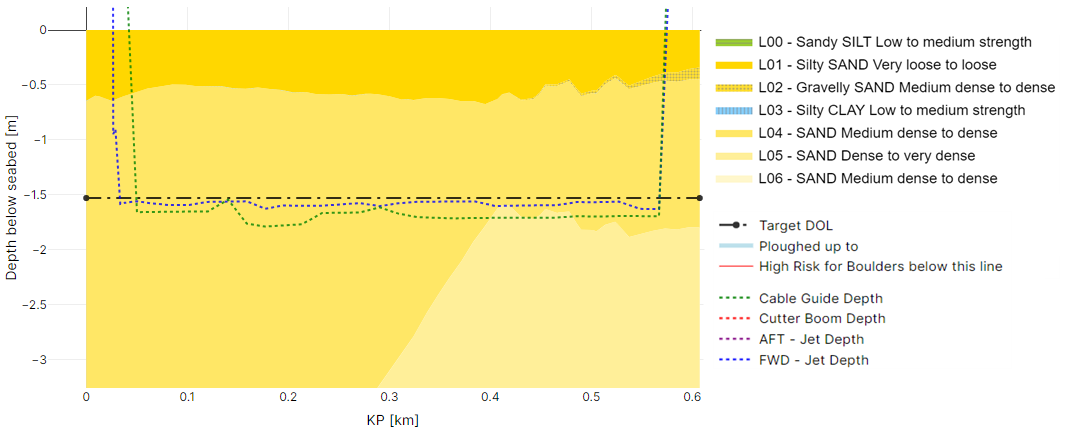}
    \caption{Soil-layer depths for a representative cable (dashboard screenshot); only layers near the ACG and jet depth are relevant.}
    \label{fig:OssK29SoilLayersDepth}
\end{figure}

\begin{figure}
    \centering
    \includegraphics[width=0.42\textwidth]{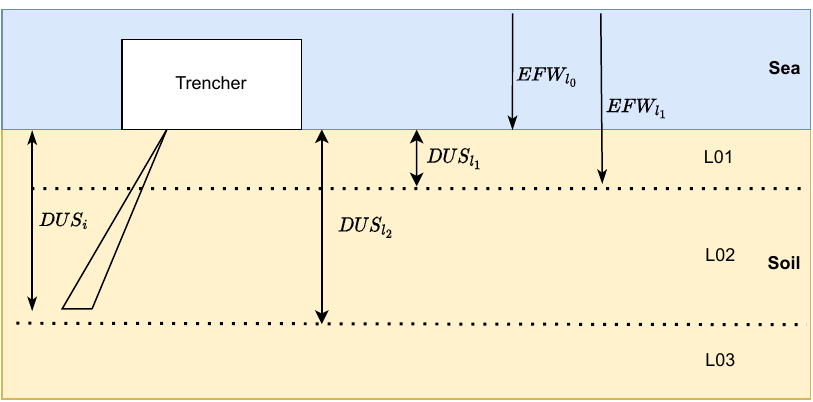}
    \caption{Soil-layer geometry and definitions. Single-tipped arrows denote elevations; double-tipped arrows denote distances.}
    \label{fig:soilLayersDiagram}
\end{figure}

\textbf{Soil-layer proportions.} Soil composition is provided as upper boundaries
of numbered layers, of which only those intersecting the trencher's tools are
relevant (Figure~\ref{fig:OssK29SoilLayersDepth}). For each layer $l_n$ we know
its elevation from the water surface; for each instrument $i$ we know its depth
under seabed $DUS_i$. Because numerically successive layers are monotonically
deeper (by relative density or cohesiveness, Eq.~\ref{eq:data:relativeDensity}),
we compute the fraction $p_{l_n,i}$ of instrument depth occupied by each layer
across three cases: layer entirely below the instrument, entirely within it, or
straddling its lower boundary (Eq.~\ref{eq:methodology:layerProportions};
Figure~\ref{fig:soilLayersDiagram}). These proportions encode soil stratigraphy
in a form the model can weigh.
\begin{equation}
\label{eq:data:relativeDensity}
    DUS_{l_0} < DUS_{l_1} < DUS_{l_2} < \dots
\end{equation}

\begin{figure*}
\begin{equation}
\label{eq:methodology:layerProportions}
p_{l_n,i} =
\begin{cases}
      0 & \text{if } DUS_{l_n} > DUS_i, \\[2pt]
      \dfrac{DUS_{l_{n+1}}-DUS_{l_n}}{DUS_i} & \text{if } DUS_{l_{n+1}} < DUS_i, \\[6pt]
      \dfrac{(DUS_{l_{n+1}} - DUS_{l_n}) - (DUS_{l_{n+1}} - DUS_i)}{DUS_i} & \text{otherwise.}
\end{cases}
\end{equation}
\end{figure*}
\textbf{Lagged aggregations and standardisation.} Because relevant signals often
manifest as windowed surges rather than instantaneous values and are delayed by
tool geometry (Eq.~\ref{eq:data:lead}), the pipeline exposes lagged windowed
aggregations as first-class, configurable features. Continuous features are
optionally Z-score standardised, $z=(x-\mu)/\sigma$, using $\mu,\sigma$ from
\emph{past} cables only to prevent leakage; interpretability-critical features
are left unstandardised.

\subsection{Labelling Harmful Load Peaks}\label{subsec:loadPeakLabel}
Because the hard retraction threshold is rarely reached and sudden surges are the
primary hazard (Section~\ref{ch:data}), we convert continuous ACG load into a
binary \emph{peak} label. Over a rolling window we compute
the mean $\mu$ (trend) and standard deviation $\sigma$ (variability) of the ACG
load (Eq.~\ref{eq:rq2:musigma}); an observation is labelled a peak when it
exceeds $\mu$ plus a multiple (calibrated by trial in the 2--3 range) of $\sigma$.
Multiple window widths capture both transient spikes and rises following stable
periods (Section~\ref{ch:results}).
\begin{equation}
\label{eq:rq2:musigma}
\begin{array}{cl}
    \mu(f_{[a,b]}) = &\frac{\sum_{i=a}^{b} f(i)}{b-a},  \\
    \sigma(f_{[a,b]}) = &\sqrt{\frac{\sum_{i=a}^{b}\big(f(i)-\mu(f_{[a,b]})\big)^2}{b-a}}. 
\end{array}
\end{equation}

\subsection{Interpretable On-Device Model}\label{sec:pipeline:model}
We deliberately use \emph{logistic regression} as the classifier. Although a
non-linear model might fit better where relationships are non-linear, logistic
regression is inherently interpretable, satisfying safety-critical sign-off and
prior-knowledge explainability requirements~\cite{beckh2021explainable} within a
kilobyte footprint suitable for on-device inference. It models peak probability
through the sigmoid (Eq.~\ref{eq:methodology:logisticFunction}); a positive
coefficient $\beta_j$ raises the log-odds of a peak as feature $X_j$ increases,
and the sign and magnitude of each coefficient provide the audit signal for
knowledge verification. Parameters are fitted by minimising average log-loss
over the training cables.
\begin{equation}
\label{eq:methodology:logisticFunction}
\hat{p}=\hat{P}(Y=1\mid X)=\frac{e^{\beta_0+\beta^{\top}X}}{1+e^{\beta_0+\beta^{\top}X}}.
\end{equation}

\section{Results and Analysis}\label{ch:results}
\subsection{Load-Peak Labelling}\label{sec:results:labels}
\begin{figure}
    \centering
    \includegraphics[width=.46\textwidth]{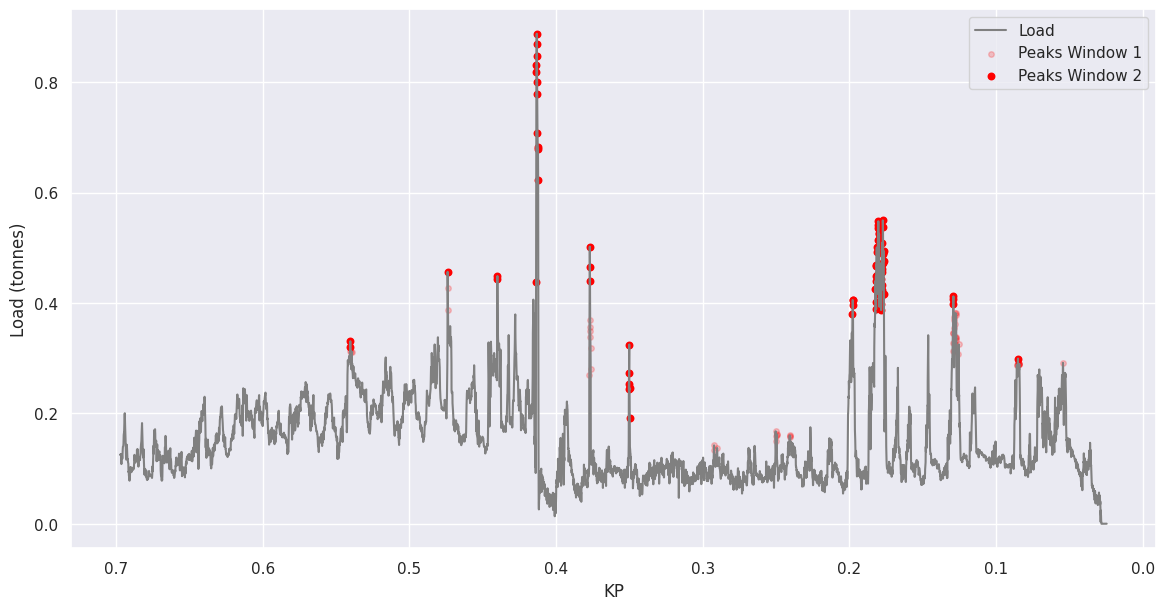}
    \caption{Load-peak labelling (1200\,s centred window) for the K25--K33 cable; the rule captures both sustained high load and abrupt surges.}
    \label{fig:peakAnalysis2output}
\end{figure}

Applying the rolling mean/standard-deviation rule
(Eq.~\ref{eq:rq2:musigma}) with three complementary windows (600\,s and 1200\,s
centred windows at $2.5\sigma$ for transient and broader spikes, plus a 30\,s
trailing window at $3\sigma$ for surges following stable periods) labels both
abrupt spikes and sustained rises. Figure~\ref{fig:peakAnalysis2output}
illustrates the combined labelling on a representative cable; full per-cable
classifications for Kaskasi are in the replication package.

\subsection{Empirical Verification of Expert Speculations}\label{sec:results:speculations}
\label{ch:rq2:sec:speculationAnalysisVerification}
Each testable speculation (Table~\ref{tab:speculationsAcgLoad}) is encoded per
Table~\ref{tab:modelEncoding} and assessed by per-cable and cumulative
coefficient sign and stability. For Kaskasi~II, speculations~1 (steering),~6
(sword depth), and~7 (boulders) are not assessable because the uniform,
boulder-free sandy soil neither contained boulder hazards nor triggered
protective steering or sword-retraction manoeuvres, so we focus on
speculations~2--5 (Table~\ref{tab:modelSpeculationEncoding}). A speculation is
\emph{confirmed} and admitted to the knowledge base only when its effect is
consistent and in the hypothesised direction.
\begin{table*}
\footnotesize
\caption{Encoded speculations and their empirical outcome (Kaskasi~II).}
\centering
    \begin{tabular}{lll}
    \toprule
    Speculation (feature)                 & Encoding (agg., window) & Outcome \\
    \midrule
    Increasing DoL ($\Delta EFS_{ACG}$)   & diff, --          & \textbf{Confirmed} \\
    Instrument-load consistency ($\sigma^2$ ACG) & var, $(-100,30)$ & Confirmed (less stable) \\
    Trencher speed (direct)               & avgdiff, $(-3,3)$ & Not supported \\
    Trencher speed (lagged)               & avgdiff, $(-60,24)$ & Not supported \\
    Soil density \& cohesiveness (layers L00--L04) & mean, $(-60,60)$  & Inconsistent \\
    \bottomrule
    \end{tabular}
    \label{tab:modelSpeculationEncoding}
\end{table*}

\begin{figure}
    \centering
    \includegraphics[width=.43\textwidth]{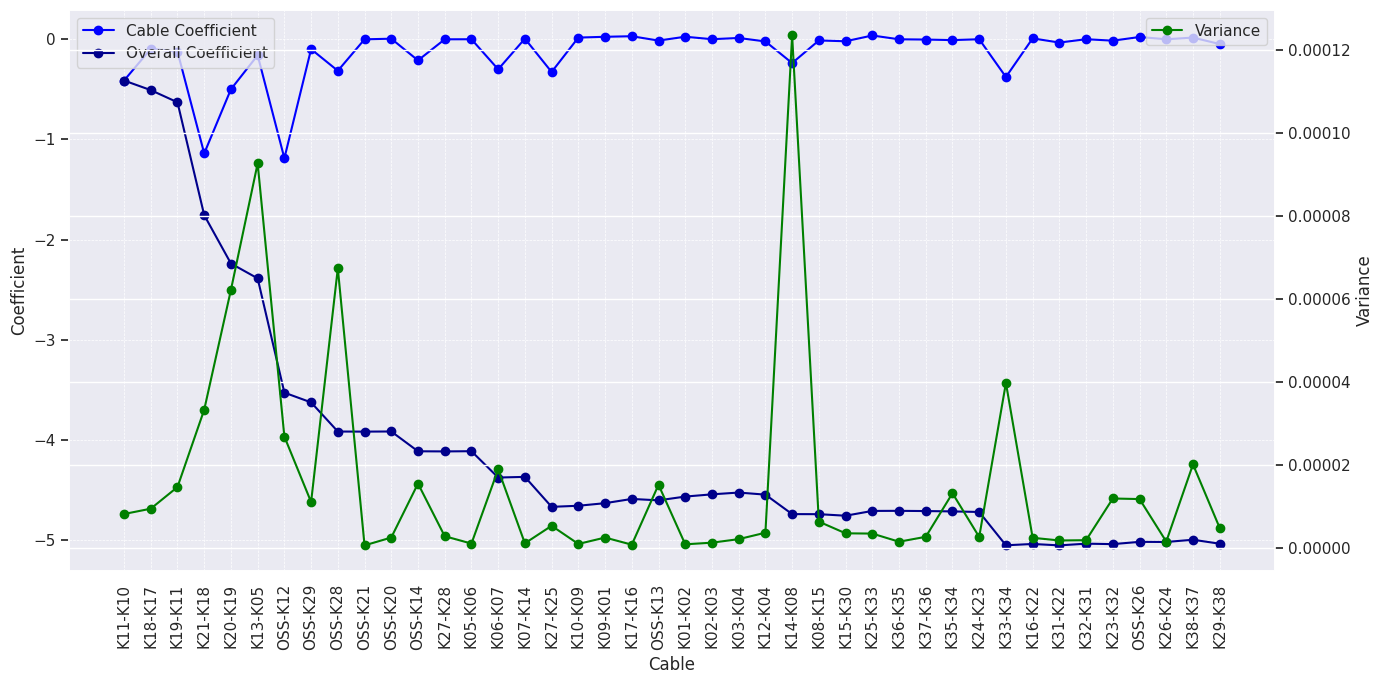}
    \caption{Fitted coefficients of $\Delta EFS_{ACG}$ on peak probability across cables (cable-specific and cumulative): a stable effect in the expected direction.}
    \label{fig:rq2_acgdepthdiff}
\end{figure}

\textbf{Increasing Depth of Lowering (confirmed).} The immediate change in ACG
elevation $\Delta EFS_{ACG}(t)=EFS_{ACG}(t)-EFS_{ACG}(t-1)$ shows a clear,
stable \emph{negative} relation to peak probability (deepening raises peak
risk), as hypothesised. Cable-specific coefficients (especially high-variance
ones) are uniformly negative and the cumulative coefficient decreases
monotonically (Figure~\ref{fig:rq2_acgdepthdiff}), so the speculation enters the
knowledge base.

\begin{figure}
    \centering
    \includegraphics[width=.43\textwidth]{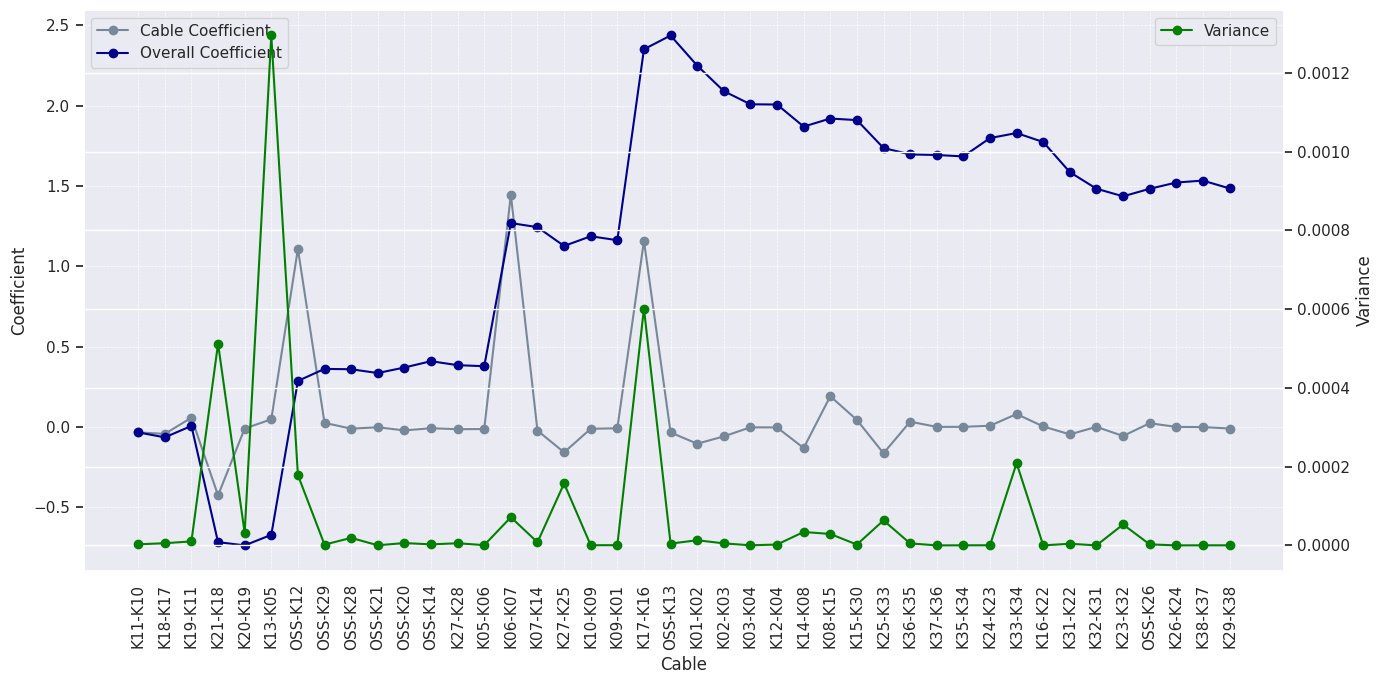}
    \caption{Fitted coefficients of ACG-load variance on peak probability: positive but less stable than DoL.}
    \label{fig:rq2_acgLoadStability}
\end{figure}

\textbf{Instrument-load consistency (confirmed, weaker).} The recent variance of
ACG load $\sigma^2(F_{ACG,[t-100,t-70]})$ has the expected positive effect, though
less consistently, driven mainly by a few cables (e.g.\ K17--K16). Because the
effect is predominantly positive it is admitted, while cases where low variance
still precedes a peak remain informative (Figure~\ref{fig:rq2_acgLoadStability}).

\begin{figure}
    \centering
    \includegraphics[width=.43\textwidth]{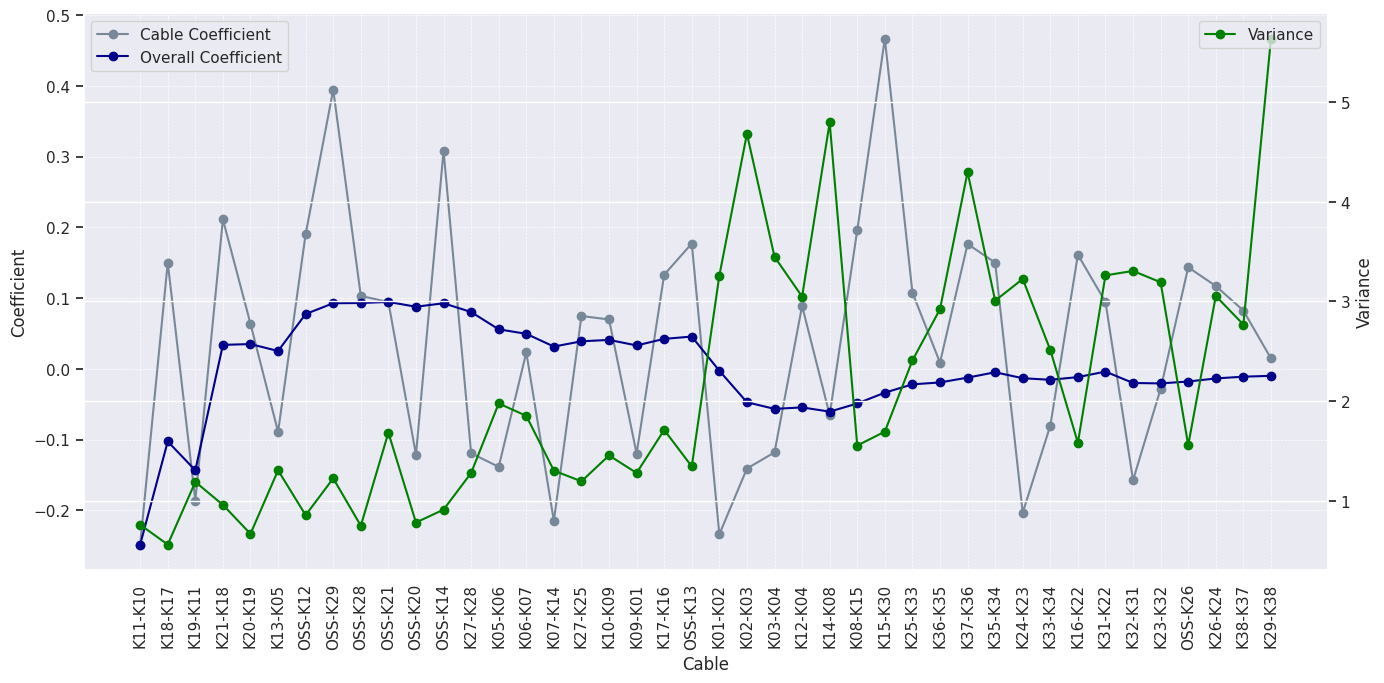}
    \caption{Fitted coefficients of short-window speed increase on peak probability: no consistent effect.}
    \label{fig:rq2_speedShortWindow}
\end{figure}

\textbf{Trencher speed (refuted).} Neither the immediate
($\mu(\Delta v_{[t-3,t]})$) nor the lagged ($\mu(\Delta v_{[t-60,t-36]})$) speed
change yields a consistent effect; both fluctuate around zero and decay as data
accumulates (Figure~\ref{fig:rq2_speedShortWindow}). These features are therefore
\emph{not} added to the knowledge base, refuting a widely held belief, though
they remain candidates for interaction effects during exploration (operators may
pre-emptively act before accelerating).

\textbf{Soil density and cohesiveness (inconsistent).} Layer proportions do not
influence peaks in a uniform direction; cohesive silt and clay layers (L00, L03)
appear to raise peak likelihood inconsistently, so they are flagged for
exploration rather than certified. Full per-layer coefficients are in the
replication package.

\subsection{Predictive Performance and Operational Benefit}\label{sec:results:performance}
With the confirmed knowledge encoded, the validated pipeline anticipates harmful
ACG-load peaks up to three minutes ahead at 0.84 AUC under rolling temporal
cross-validation within a 32\,kB footprint
(Table~\ref{tab:results:performance}). An ablation against an otherwise identical
\emph{knowledge-free} baseline (raw features, no encoded speculations) shows that
injecting prior knowledge roughly \emph{halves the false-alarm rate} at matched
recall, confirming that knowledge engineering, rather than the model class alone,
drives the gain. Replaying the model's speed recommendations against the recorded
campaigns indicates an 11\% reduction in non-productive time.
\begin{table}
\caption{Validated pipeline vs.\ knowledge-free baseline (rolling temporal CV; false-alarm rate at matched recall).}
\centering
\footnotesize
\setlength{\tabcolsep}{4pt}
\begin{tabular}{@{}p{3.0cm}cc@{}}
\toprule
Metric & Knowledge-free & Knowledge-informed \\
\midrule
AUC (peak anticipation)     & lower    & \textbf{0.84} \\
False-alarm rate            & baseline & $\approx$\,50\% lower \\
Lead time                   & --       & up to 3\,min \\
On-device footprint         & 32\,kB   & 32\,kB \\
Implied non-productive time & --       & $-11$\% \\
\bottomrule
\end{tabular}
\label{tab:results:performance}
\end{table}

\subsection{Out-of-Sample Validation: South Fork}\label{ch:results:southfork}
To probe external validity (RQ3) we re-apply the \emph{certified} pipeline,
unchanged, to South Fork Wind, a more recent campaign with harder, more variable
soil and a stronger productivity emphasis. We first re-test the confirmed
speculations, then exploit the richer soil to explore new effects; per-cable
burial intervals and soil identifiers are in the replication package.

\begin{figure}
    \centering
    \includegraphics[width=.43\textwidth]{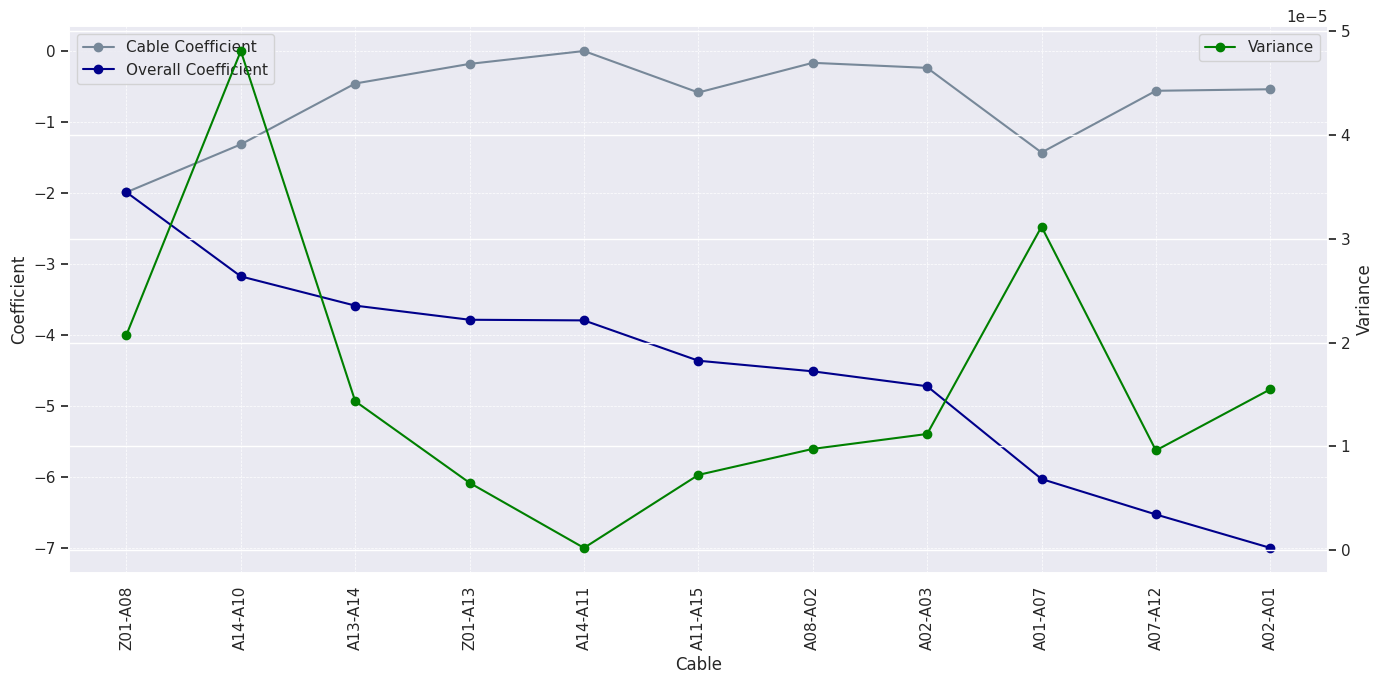}
    \caption{Fitted coefficients of $\Delta EFS_{ACG}$ on peak probability for South Fork: the confirmed effect transfers out-of-sample.}
    \label{fig:rq2_acgdepthdiff_southfork}
\end{figure}

\textbf{Confirmed speculations transfer and strengthen.} The Depth-of-Lowering
effect ($\Delta EFS_{ACG}$) remains stable and consistently strengthens in the
expected direction (Figure~\ref{fig:rq2_acgdepthdiff_southfork}), and the
instrument-load-consistency effect is \emph{more} consistent than on Kaskasi.
Both validated pieces of knowledge thus generalise to a materially different
campaign.
\textbf{Boulder risk is not a direct cause.} With boulder data available here, we
test speculation~7. The average boulder proportion over the ACG,
$\mu(p_b,{[t-60,t]})$, shows no consistent effect and decays over time: boulder
risk \emph{alone} does not explain peaks.

\begin{figure}
    \centering
    \includegraphics[width=.43\textwidth]{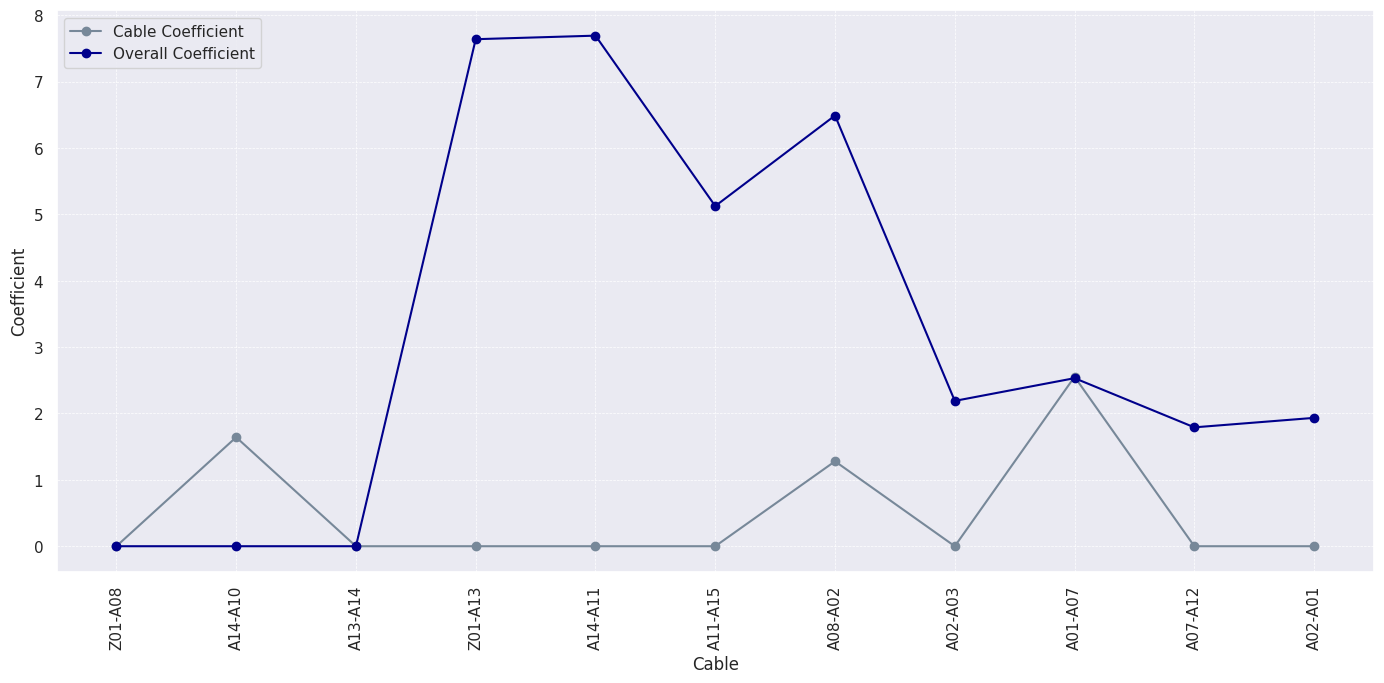}
    \caption{Fitted coefficients of the boulder-probability\,$\times$\,L01-proportion interaction on peak probability (South Fork): consistent in the expected direction.}
    \label{fig:southForkL02Boulder}
\end{figure}

\textbf{Exploration surfaces an interaction.} The harder soil yields more feature
variance and thus deeper analysis. Adding layer proportions one at a time (to
manage their inherent collinearity, since proportions sum to one) reveals that a
seemingly strong negative effect of a very dense layer (L02) is a \emph{spurious}
correlation once the cable is examined directly, underscoring the value of
pipeline auditability. By contrast, the \emph{interaction} between boulder
probability and the denser L01 layer is consistent in the expected direction
(Figure~\ref{fig:southForkL02Boulder}): peaks become more likely when high
boulder risk coincides with denser soil over the ACG. Decomposing a high-effect
cable into trenching phases shows complexity and productivity tracking these
combined conditions, yielding concrete (if tentative) speed recommendations that
the TinyMLOps solution could enact live.
Overall, the pipeline generalises to South Fork and, aided by the campaign's
greater variance, yields deeper insight while confirming RQ3.

\section{Discussion}\label{ch:discussion}
\subsection{Revisiting the Research Questions}

\begin{researchqbox}
\noindent\textbf{Answer to RQ1 (Engineering).}
Decoupling costly per-cable data construction from lightweight model fitting, and
expressing domain hypotheses as YAML-configured lagged aggregations over a
regularised logistic classifier, yields an auditable TinyMLOps pipeline fitting
within a 32\,kB Cortex-M4 budget and removing the 15-minute dashboard latency.
\end{researchqbox}

\begin{researchqbox}
\noindent\textbf{Answer to RQ2 (Efficacy).}
Cross-cable coefficient-stability analysis separates durable physical effects
(deepening the ACG and elevated load variance) from unsupported operator beliefs
(marginal speed increases). Admitting only certified knowledge halves false
alarms at 0.84 AUC up to three minutes ahead and implies an 11\% cut in
non-productive time.
\end{researchqbox}

\begin{researchqbox}
\noindent\textbf{Answer to RQ3 (Generalisability).}
Applied unchanged to South Fork Wind's harder, boulder-prone soils, the pipeline
reproduces and strengthens the certified Kaskasi effects while surfacing a
meaningful interaction (boulder risk\,$\times$\,gravelly-sand layer L01) and
exposing a spurious single-layer correlation (L02). The knowledge-verification
architecture transfers across campaigns even where soil regimes differ.
\end{researchqbox}

Together, the three RQs expose a key socio-technical dynamic of
human-in-the-loop CPS: \emph{operator compensation can mask causal risk in
observational logs}. Although experts expected acceleration to trigger surges,
crews accelerate selectively under benign conditions or adjust jetting pressures
pre-emptively, flattening the marginal $\Delta v$ correlation. Surfacing
per-cable coefficients alongside feature variance prevented conflating control
policy with plant physics while retaining speed for interaction tests.

\subsection{Implications for Software-Engineering Practice}
\textbf{Knowledge-as-configuration as a socio-technical boundary object.}
Encoding expert speculation as YAML-described, lagged feature transformations
(Table~\ref{tab:modelEncoding}) decouples \emph{what} specialists hypothesise
from \emph{how} the pipeline executes. The configuration file acted as a
version-controlled interface between marine/geotechnical and ML engineers,
letting specialists revise lag windows without touching code while the two-mode
engine enforced a promotion gate from \emph{exploration} to \emph{validation}.
In knowledge-intensive CPS, declarative hypothesis specifications offer higher
leverage than cloud-style model registries.

\textbf{Explainability as a confound debugger, not a performance tax.}
Choosing a linear log-odds model and plotting per-cable versus cumulative
coefficients enabled both safety sign-off and empirical debugging. On South Fork,
inspecting the evidence chain revealed that a strong negative coefficient on very
dense sand (L02) was an artefact of collinearity and phase boundaries rather than
physical relief, preventing a hazardous rule from reaching operators. An opaque
neural model would have absorbed that spurious correlation silently; here,
interpretability and false-alarm reduction reinforced one another.

\textbf{Evaluation under both environmental drift and adaptive control.}
Streaming CPS data violates i.i.d.\ assumptions twice: seabed stratigraphy shifts
across cables, and operators intervene adaptively within each pass. Pairing
rolling temporal cross-validation (with past-only standardisation) and cross-cable
coefficient-stability checks guards against look-ahead leakage and single-cable
interventions, and should be standard for streaming CPS ML pipelines.

\textbf{Kilobyte budgets dictate a split TinyMLOps topology.} Targeting a 32\,kB
Cortex-M4 deployment shaped the end-to-end architecture rather than compressing a
model post hoc. Heavy governance tasks (tag discovery, depth/elevation
normalisation, KP-monotonic windowing, and stratigraphic geometry) run once
upstream during data construction, leaving only compact rolling statistics and
linear scoring for on-device MicroPython execution~\cite{faubel2023mlops}.

\textbf{Environmental and software sustainability.} Frugal model selection curbs
both software and operational carbon footprints~\cite{tamburri2020sustainable}:
regularised linear training on 4.4\,GB of telemetry takes seconds on a CPU, and
milliwatt Cortex-M4 inference avoids GPU clusters and continuous vessel-to-shore
streaming. Operationally, cutting non-productive trenching time and re-burials by
11\% directly reduces offshore support-vessel hours and marine-fuel emissions.

\subsection{Implications for Research and Reproducibility}
Our findings refine taxonomies of informed
ML~\cite{von2021informed,beckh2021explainable} by showing that \emph{prior
knowledge requires differentiated engineering}: deterministic physical priors
(Eqs.~\ref{eq:data:lead}--\ref{eq:methodology:layerProportions}) belong in data
construction, whereas fallible expert heuristics
(Table~\ref{tab:speculationsAcgLoad}) require an empirical falsification gate.
Because semantic drift across 650 tags and coarse DPR logs can corrupt models
upstream, releasing KP-window extraction, YAML encodings, and per-cable stability
plots keeps confirmed and refuted effects auditable despite proprietary raw logs.

\section{Threats to Validity}\label{ch:threats}
\subsection{Construct Validity}
Our central construct, a \emph{harmful} ACG-load peak, lacks a domain-standard
definition; we operationalise it via a rolling mean/standard-deviation rule whose
multiplier (2--3$\sigma$) and window widths were calibrated by trial. A different
operationalisation could shift which events count as peaks. We mitigate this by
using multiple windows capturing both transient and sustained surges, grounding
the rule in operators' concern for sudden spikes, and releasing the labelling
code for scrutiny. Likewise, the 11\% non-productive-time reduction is estimated
by \emph{replaying} recommendations against recorded campaigns rather than a live
trial, and should be read as an indicative estimate. Finally, speculations were
elicited through individual expert discussions; structured focus groups would
strengthen construct coverage.

\subsection{Internal Validity}
The speculation-verification analysis makes directional claims that could be
confounded. Soil-layer proportions are strongly collinear (summing to one), so
we add them one at a time and inspect the evidence chain rather than reading
coefficients in isolation, which exposed a \emph{spurious} dense-layer
effect on South Fork. Human behaviour (steering, operator reaction time) and
cross-cable concept drift are further confounders, mitigated by judging each
effect on its cross-cable stability and cumulative trend under leak-free rolling
temporal cross-validation. Unstable or single-cable-driven effects are treated as
exploratory, not confirmed.

\subsection{External Validity}
The study covers one industrial partner, one trenching ROV, and two offshore-wind
campaigns, modelling first-pass trenching only; Kaskasi~II is predominantly sandy,
which may understate harder-soil effects. Generalisation beyond subsea
trenching to other CPS classes is argued by construction: the inter-array cables
form a varied-length sample, and South Fork provides out-of-sample validation
under materially harder conditions where the \emph{method}, not merely the
trained model, transferred. Accordingly, we claim generalisability for the
\emph{pipeline pattern} rather than specific fitted coefficients.

\subsection{Conclusion Validity}
Our conclusions rest on coefficient sign and stability across a finite set of
cables (42 for Kaskasi, 11 for South Fork) rather than formal hypothesis tests
with effect sizes, and AUC is a single operating summary. Trial-and-error
calibration of the labelling rule also risks experimenter bias. To guard against
reporting anti-patterns, we \emph{pre-specified} expert speculations before
fitting (avoiding HARKing) and report \emph{refuted} (trencher speed) and
inconsistent (soil-layer) effects alongside confirmations. Releasing the complete pipeline code, YAML configurations, and supplementary
evaluation outputs enables independent replication and stricter inferential
testing on analogous CPS telemetry.

\section{Conclusion and Future Work}\label{ch:conclusion}
This field study showed that an explainable, knowledge-centred TinyMLOps
pipeline can be engineered, deployed on a kilobyte-scale embedded target, and
operated as trustworthy decision support inside a safety-critical CPS. Across two
offshore-wind campaigns, encoding expert beliefs as testable features halved
false alarms, confirmed a counter-intuitive cause, refuted a widely held one, and
transferred out-of-sample within a 32\,kB footprint.

\textbf{Lessons learned.} Three lessons generalise beyond the case: (1)~in
knowledge-intensive CPS, a \emph{declarative, testable encoding of domain
knowledge} is a high-leverage MLOps interface; (2)~\emph{explainability and
resource frugality reinforce each other} through auditable, low-parameter models;
and (3)~\emph{evaluation must assume drift} via rolling temporal
cross-validation.

\textbf{Future work.} On \emph{operations}, deployment should be hardened with
concept-drift detection, automated retraining triggers, hardware-degradation
guards, and human--computer interaction design for the operator interface,
closing the MLOps loop. On \emph{method}, speculation elicitation should be
formalised (e.g.\ focus groups), the labelling rule grounded in a dedicated
study, and lagged windows adapted dynamically to trenching speed. On
\emph{modelling}, explainable non-linear models could be explored where linear
effects underfit while preserving the audit chain. On \emph{evidence}, a
controlled field trial should measure the replay-estimated productivity gain live,
and replications across other Industry~4.0 CPS classes should test how far the
pipeline pattern generalises.

\section{Data Availability}\label{sec:replication}
The \texttt{trencher} Python package, SQL queries, declarative pipeline
configuration (\texttt{config.yaml}), and supplementary campaign metadata and
evaluation plots are archived under the MIT licence on Zenodo
(\url{https://doi.org/10.5281/zenodo.23614280}). Raw sensor logs, DPRs, and
geotechnical survey tables remain proprietary to the industrial partner under
commercial confidentiality constraints; researchers with analogous CPS telemetry
can adapt and run the pipeline via \texttt{config.yaml}.

\section*{Acknowledgments}
The authors thank the engineers of the industrial partner for the expert
elicitation and operational data access.

\bibliographystyle{IEEEtranN}
\bibliography{references}


\end{document}